\documentclass[aps,prd,twocolumn,superscriptaddress,nofootinbib]{revtex4-2}

\usepackage{supertabular} 
\usepackage{placeins}
\usepackage{epsfig}
\usepackage{graphicx}

\usepackage{float}
\usepackage{mathtools}
\usepackage{array}
\usepackage{dcolumn}

\usepackage{latexsym}
\usepackage{amsmath}
\usepackage{amssymb}
\usepackage{amsfonts}
\usepackage{bm}

\usepackage{color}
\definecolor{purple}{rgb}{0.5,0,0.5}
\definecolor{blue}{rgb}{0.0,0,0.9}
\definecolor{prdblue}{rgb}{0.133,0.118,0.498}
\usepackage[colorlinks=true, pdfstartview=FitV, linkcolor=prdblue, citecolor= prdblue, urlcolor=prdblue]{hyperref}

\def\tstrut{\vrule height3.25ex depth0pt width0pt} 

\begin{document}


\title{Electromagnetic form factors of the $\mathbf{\Omega}$ baryon: space-like structure and time-like extrapolation}

\author{L. Albino}
\email[]{luis.albino.fernandez@gmail.com}
\affiliation{Departamento de F\'isica, Universidad de Sonora, Boulevard Luis Encinas J. y Rosales, 83000, Hermosillo, Sonora, Mexico.}

\author{B. Almeida-Zamora}
\email[]{bilgai\_almeidaz@hotmail.com}
\affiliation{Departamento de Sistemas F\'isicos, Qu\'imicos y Naturales, Universidad Pablo de Olavide, Avenida Rectora Rosario Valpuesta 1, 41089, Dos Hermanas, Sevilla, Spain.}

\author{A. Bashir}
\email[]{adnan.bashir@umich.mx}
\affiliation{Departamento de Ciencias Integradas, Universidad de Huelva, E-21071 Huelva, Spain.}
\affiliation{Facultad de Ingenier\'ia, Universidad Aut\'onoma de Quer\'etaro, Quer\'etaro, Quer\'etaro 76010, M\'exico}

\author{J.J. Cobos-Mart\'inez}
\email[]{jesus.cobos@unison.mx}
\affiliation{Departamento de F\'isica, Universidad de Sonora, Boulevard Luis Encinas J. y Rosales, 83000, Hermosillo, Sonora, Mexico.}

\author{K. Raya}
\email[]{khepani.raya@dci.uhu.es}
\affiliation{Departamento de Ciencias Integradas, Universidad de Huelva, E-21071 Huelva, Spain.}

\author{J.~Segovia}
\email[]{jsegovia@upo.es}
\affiliation{Departamento de Sistemas F\'isicos, Qu\'imicos y Naturales, Universidad Pablo de Olavide, Avenida Rectora Rosario Valpuesta 1, 41089, Dos Hermanas, Sevilla, Spain.}

\date{\today}

\begin{abstract}
We calculate the elastic electromagnetic multipole form factors of the $\Omega^-$ baryon using a symmetry-preserving vector$\,\otimes\,$vector contact interaction in rainbow--ladder truncation and a Poincar\'e-covariant quark--diquark Faddeev equation. A conserved electromagnetic current yields the electric monopole, magnetic dipole, electric quadrupole, and magnetic octupole form factors. We examine their sensitivity to a phenomenological transverse dressed-quark anomalous magnetic moment (AMM). The magnetic and higher multipoles show the strongest dependence, whereas the AMM contribution is progressively suppressed with increasing momentum transfer. The calculated magnetic moment is larger in magnitude than the experimental value, and a positive AMM contribution increases this discrepancy, exposing a limitation of the present truncation. We also construct a model-dependent time-like extrapolation of the space-like results, motivated by their asymptotic behavior, and use it to evaluate the effective form factor measured in $e^+e^-\to\Omega^-\bar{\Omega}^+$. The extrapolation describes the broad trend at larger energies but does not incorporate resonances, final-state interactions, unitarity cuts, or the complex phase structure required near threshold.
\end{abstract}


\maketitle


\section{INTRODUCTION}
\label{sec:introduction}

Understanding the internal structure of hadrons in terms of quark and gluon degrees of freedom remains a central goal of strong-interaction physics~\cite{AbdulKhalek:2021gbh, Accardi:2023chb, Ai:2025xop}. Among the most powerful probes of hadron structure are electromagnetic form factors, which encode information about the spatial distributions of charge and magnetization and provide direct insight into the nonperturbative dynamics responsible for binding quarks and gluons into observable matter~\cite{Perdrisat:2006hj, Pacetti:2014jai, Punjabi:2015bba}. Over the past decades, extensive experimental and theoretical efforts have established a detailed picture of the electromagnetic structure of the nucleon and its excited states~\cite{Kelly:2004hm, Arrington:2007ux, Ye:2017gyb, Tiator:2011pw, Ronchen:2012eg, Aznauryan:2012ba, Kamano:2013iva, Achenbach:2023pba, Ramalho:2023hqd, ParticleDataGroup:2024cfk}. In contrast, considerably less is known about the electromagnetic properties of baryons composed predominantly of strange quarks.

The $\Omega$ baryon, a spin-$3/2$ state composed exclusively of strange valence quarks, occupies a distinctive position in the baryon spectrum~\cite{Capstick:1986ter, Hyodo:2020czb}. The absence of light valence quarks reduces its sensitivity to long-range pion-cloud effects and makes its electromagnetic structure a useful setting in which to study the role of the strange-quark mass, SU(3)-flavor symmetry, and dynamical chiral symmetry breaking~\cite{Eichmann:2016yit, Qin:2019hgk}. Its stability under the strong interaction also makes it suitable for lattice-QCD studies and future experiments. Nevertheless, empirical information remains scarce, especially in the time-like region~\cite{Ramalho:2020laj}. Existing theoretical results comprise lattice-QCD simulations in the space-like domain~\cite{Boinepalli:2009sq,Alexandrou:2010jv}, a momentum-dependent covariant three-body continuum calculation~\cite{Sanchis-Alepuz:2013iia}, and phenomenological analyses and extrapolations into the time-like domain~\cite{Ramalho:2020laj, Fu:2025vkq}.

Continuum approaches to QCD, such as those based on Dyson--Schwinger equations (DSEs), provide a Poincaré-covariant framework capable of describing both mesons and baryons on equal footing~\cite{Roberts:1994dr, Alkofer:2000wg, Maris:2003vk, Cloet:2013jya}. When combined with symmetry-preserving truncations~\cite{Eichmann:2008ae, Binosi:2016rxz, Qin:2019oar}, most notably the rainbow--ladder (RL) approximation, these methods have achieved considerable success in describing hadron masses, decay constants, and form factors~\cite{Eichmann:2008ef, Qin:2011xq, Chang:2013nia, Xu:2015kta, Raya:2015gva, Lu:2017cln, Chen:2018nsg, Lu:2019bjs, Cui:2020rmu, Chen:2020wuq, Qin:2020rad, Chen:2021guo}. In the baryon sector, the covariant Faddeev equation furnishes a natural description of three-quark bound states, incorporating diquark correlations that emerge dynamically from the underlying quark--quark interaction~\cite{Chen:2017pse, Chen:2019fzn, Barabanov:2020jvn, Liu:2022ndb, Liu:2022nku}.

Continuum calculations span a range of truncations, from momentum-dependent kernels solved in a covariant three-body framework to contact interactions formulated in a quark--diquark picture. The former retain a more complete momentum dependence; the latter provide an algebraically tractable realization that preserves the relevant symmetry constraints. The present calculation is not intended to supersede the momentum-dependent three-body result of Ref.~\cite{Sanchis-Alepuz:2013iia}. Instead, it addresses a narrower question within a fixed contact-interaction framework: how the four $\Omega$ multipoles respond to a phenomenological transverse dressed-quark AMM. This comparison exposes which observables are most sensitive to that vertex component and provides a quantitative assessment of the limitations of the adopted truncation.

In this work, we employ a vector$\,\otimes\,$vector contact interaction (SCI) within the RL truncation of the quark DSE and baryon Faddeev equations in order to compute the electromagnetic form factors of the $\Omega$ baryon. This framework preserves the relevant symmetries of QCD while allowing for a tractable, algebraic treatment of the bound-state problem~\cite{Roberts:2011wy, Chen:2012qr, Segovia:2013rca, Yin:2019bxe, Yin:2021uom}. The present study extends earlier contact-interaction investigations of baryon structure~\cite{Wilson:2011aa, Segovia:2014aza, Raya:2021pyr, Albino:2025fcp, Albino:2025bnr} by calculating all four elastic $\Omega$ multipoles in the space-like domain and examining their sensitivity to the AMM parameter. We additionally explore a model-dependent continuation of the resulting parametrizations into the time-like domain. This continuation is used to construct the effective form factor and is not a dynamical calculation of the physical time-like analytic structure.

The electromagnetic current is constructed to be conserved within the adopted quark--diquark framework. Its transverse quark--photon component contains both vector dressing and the phenomenological AMM term~\cite{Ward:1950xp, Takahashi:1957xn, Ball:1980ay}. For a spin-$3/2$ particle, the current is characterized by four independent form factors, corresponding to the electric monopole, magnetic dipole, electric quadrupole, and magnetic octupole moments~\cite{Nozawa:1990gt}. We calculate these form factors in the space-like region and subsequently examine a time-like extrapolation motivated by their asymptotic behavior~\cite{Maris:1999bh, Eichmann:2011vu, Segovia:2015hra, Ramalho:2020laj}.

This article is organized as follows. In Sec.~\ref{sec:theory} we briefly summarize the theoretical framework, including the Faddeev equation for the dressed-quark core and the construction of the electromagnetic current. Section~\ref{sec:results} presents our results for the $\Omega$ elastic form factors in both the space-like and time-like regions. Finally, Sec.~\ref{sec:summary} contains a summary and outlook.


\section{THEORETICAL FORMALISM}
\label{sec:theory} 

\subsection{Symmetry-preserving contact interaction (SCI)}
\label{subsec:SCI}

\begin{figure}[!t]
\centerline{%
\includegraphics[clip,width=0.45\textwidth, height=0.09\textheight]{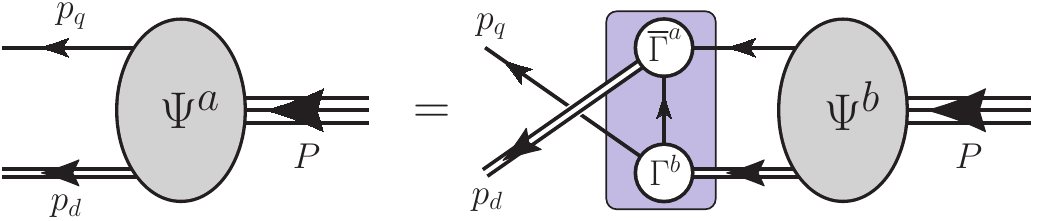}}
\caption{\label{fig:Faddeev} Poincar\'e-covariant Faddeev equation.  $\Psi$ is the Faddeev amplitude for a baryon of total momentum $P= p_q + p_d$.  The shaded rectangle demarcates the kernel of the Faddeev equation: \emph{single line}, dressed-quark propagator; $\Gamma$,  diquark correlation amplitude; and \emph{double line}, diquark propagator.}
\end{figure}

We describe baryon bound states using a Poincaré-covariant Faddeev equation, illustrated in Fig.~\ref{fig:Faddeev}. Its essential ingredients are the dressed-quark and diquark propagators, together with the diquark Bethe--Salpeter amplitudes. These elements are fully specified once the quark--quark interaction kernel is chosen.

As noted in the Introduction, we employ a symmetry-preserving regularization of a vector$\,\otimes\,$vector contact interaction, characterized by a momentum-independent gluon propagator:
\begin{eqnarray}
g^2 D_{\mu\nu} \left( p-q \right) = \frac{4 \pi \alpha_{\text{IR}}}{m_g^2} \, \delta_{\mu\nu} \,,
\label{eq:GluonPropagator}
\end{eqnarray}
where the effective gluon mass, $m_g = 0.5$ GeV, is consistent with the infrared mass scale associated with a finite gluon propagator in lattice-regularized QCD~\cite{Bowman:2004jm, Bogolubsky:2009dc, Aguilar:2008xm}. The parameter $\alpha_{\text{IR}} = 0.36\pi$ defines an effective coupling whose value is commensurate with contemporary determinations of the QCD running coupling at vanishing momentum~\cite{Binosi:2016nme, Cui:2019dwv}.

We embed Eq.~\eqref{eq:GluonPropagator} within the rainbow--ladder (RL) truncation of the Dyson--Schwinger equations, which employs a tree-level quark--gluon vertex, 
\begin{equation}
\Gamma_{\mu}^a (q,p) = \frac{\lambda^a}{2}\gamma_{\mu} \,,
\end{equation}
with $\lambda^a$ the Gell-Mann matrices. In this truncation, the dressed-quark propagator assumes the form
\begin{eqnarray}
\label{Dressed-quark propagator}
S^{-1}(p) = i \gamma \cdot p + M \,,
\end{eqnarray}
where the mass function reduces to a momentum-independent dressed-quark mass $M$, determined by the gap equation:
\begin{eqnarray}
\label{non-regularized gap equation}
M = m_q + M \frac{4 \alpha_{\text{IR}}}{3 \pi m_g^2} \int_{0}^{\infty} ds \, \frac{s}{s + M^2} \,,
\end{eqnarray}
with $m_q$ the current-quark mass. Equation~\eqref{non-regularized gap equation} is quadratically divergent and must therefore be regularized in a Poincaré-covariant manner. We employ a proper-time regularization scheme~\cite{Ebert:1996vx}, which introduces an infrared cutoff, $\tau_{\text{ir}}$, implementing confinement~\cite{Krein:1990sf, Chang:2011vu, Roberts:2012sv}, and an ultraviolet cutoff, $\tau_{\text{uv}}$, which sets the overall mass scale~\cite{Gutierrez-Guerrero:2010waf}. Accordingly, the integrand denominator in Eq.~\eqref{non-regularized gap equation} is modified as
\begin{eqnarray}
\frac{1}{s + M^2} \to \frac{e^{- ( s + M^2 ) \tau_{\text{uv}}^2} - e^{- ( s + M^2 ) \tau_{\text{ir}}^2}}{s + M^2} \,.
\label{proper-time regularization}
\end{eqnarray}
Substituting Eq.~\eqref{proper-time regularization} into the gap equation yields a finite integral equation for $M$, whose solution is nonzero even in the chiral limit ($m_q=0$), thereby realizing dynamical chiral symmetry breaking.

In this work, we adopt the parameter set of Ref.~\cite{Yin:2021uom}, namely
\begin{equation}
\frac{1}{\tau_{\text{ir}}} = 0.240\,\text{GeV} \,, \quad\quad \frac{1}{\tau_{\text{uv}}} = 0.905\,\text{GeV} \,,
\end{equation}
fixed to reproduce the pion's mass and decay constant. This choice yields a constituent strange-quark mass $M=0.53$ GeV, obtained with a current quark mass $m_q=0.17\,\text{GeV}$.


\subsection{The $\mathbf{\Omega}$ Faddeev equation}
\label{subsec:Faddeev}

Within the quark--diquark picture, the baryon's Faddeev amplitude can be expressed as a sum over contributions in which each quark is treated as a spectator in turn:
\begin{eqnarray}
\Psi = \Psi_1 +\Psi_2 +\Psi_3 \,,
\end{eqnarray}
where the subscript labels the spectator quark, and the remaining components are obtained by cyclic permutation of the quark indices.

Studies based on QCD-like interaction kernels~\cite{Chen:2019fzn} have shown that the $\Omega$ baryon is mostly dominated by axial-vector diquark correlations with flavor content $\{ss\}_{1^+}$. Accordingly, we retain only this dominant correlation and employ a simplified yet realistic representation of the $\Omega$ Faddeev amplitude:
\begin{eqnarray}
\label{Omega FA}
\Psi_3 = \Gamma^{1^{+}}_{\alpha}(p_1,p_2)\Delta^{1^{+}}_{\alpha\beta}(K)\mathcal{D}_{\beta\rho}(P)u_{\rho}(P) \,,
\end{eqnarray}
where $u_{\rho}(P)$ is the Rarita--Schwinger spinor describing an on-shell spin-$3/2$ baryon with total momentum $P$, and $K=p_1+p_2$ is the diquark momentum.

The amplitude $\Gamma_\alpha^{1^{+}}(p_1,p_2)$ denotes the canonically normalized Bethe--Salpeter amplitude of the axial-vector diquark with flavor content $\left\{ ss \right\}_{1^+}$, while the corresponding propagator is given by
\begin{eqnarray}
\Delta^{1^{+}}_{\mu \nu} (K)  &=& \left[ \delta_{\mu \nu} + \frac{K_{\mu} K_{\nu}}{m^2_{1^{+}}} \right] \frac{1}{K^2 + m^2_{1^{+}}} \,, \label{Diquark-propagator: axial}
\end{eqnarray}
where $m_{1^{+}}$ is the mass-scale associated with the corresponding diquark correlation. In this work, we use the value obtained in Ref.~\cite{Yin:2021uom}:
\begin{eqnarray}
m_{\left\{ ss \right\}_{1^+}} =  1.26 \, \hbox{GeV} \,.
\end{eqnarray}

The Dirac structure describing the quark--diquark relative momentum correlation in the $\Omega$ baryon, $D_{\beta\rho}(P)$, is considerably simplified within the contact-interaction framework, reducing to
\begin{eqnarray}
\mathcal{D}_{\beta\rho}(P) &=& d^{\{ss\}_{1^{+}}} \, \mathbf{I}_{\mathrm{D}} \, \delta_{\beta\rho} \,, \label{Faddeev Amplitude}
\end{eqnarray}
with $\mathbf{I}_{\mathrm{D}}$ defining the $4 \times 4$ identity matrix in Dirac space and $d^{\{ss\}_{1^{+}}}=1$.

As illustrated in Fig.~\ref{fig:Faddeev}, the Faddeev kernel (shaded rectangle) involves diquark breakup and reformation via exchange of a dressed-quark. We now follow Ref.~\cite{Roberts:2011cf} and implement a variant of the so-called static approximation~\cite{Buck:1992wz,Xu:2015kta}; namely, in the Faddeev equation for a baryon of type $B$, the quark exchanged between the diquarks is represented as
\begin{equation}
S_f^T=\frac{g_B^2}{M_f} \,, \label{g_B}
\end{equation}
where $f = u(d), s, c, b$ is the quark's flavor, $M_f$ is the corresponding dressed-quark mass, and the superscript $T$ denotes matrix transpose. The parameter $g_B$ is an effective coupling that encodes the strength of quark exchange in the baryon; for the $\Omega$ baryon its value is fixed phenomenologically, $g_\Omega=1.56$~\cite{Yin:2021uom}. This variant of the static approximation, originally introduced in Ref.~\cite{Buck:1992wz}, has subsequently been used in studies of a wide range of baryon properties~\cite{Yin:2019bxe, Yin:2021uom}. 

All together, the resulting Faddeev equation yields a mass for the $\Omega$ baryon:
\begin{eqnarray}
m_{\Omega} &=&  1.76 \, \hbox{GeV} \,,
\end{eqnarray}
which is slightly higher but compatible with the empirical mass~\cite{ParticleDataGroup:2024cfk}, $m_{\Omega}^e=1.67\,\text{GeV}$, and the lattice-QCD determinations~\cite{BMW:2008jgk, Brown:2014ena, Mathur:2018epb}, $m_{\Omega}^l=1.67\,\text{GeV}$, noting that the SCI approach does not incorporate meson-cloud contributions to hadron masses~\cite{Yin:2019bxe, Yin:2021uom}; for instance, it yields a bare nucleon mass of $1.14\,\text{GeV}$.


\subsection{The photon-baryon coupling}
\label{subsec:EFFs}

\begin{figure}[!t]
\centering
\includegraphics[scale=0.25]{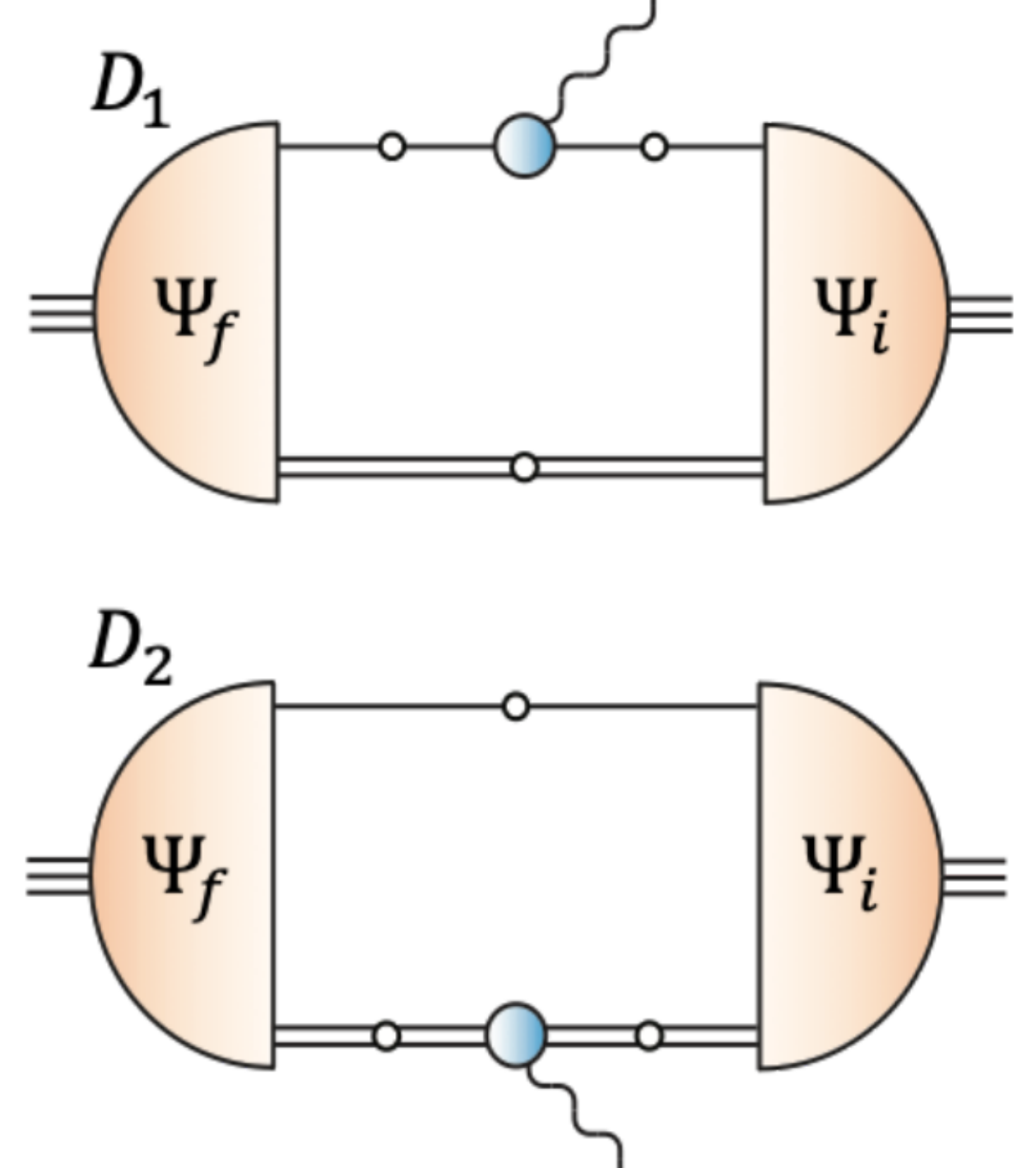}
\caption{\label{fig:EMInteractions} Diagrammatic representation of contributions for elastic EM currents in the quark-diquark picture. Dressed-quark and diquark propagators are represented by single and double lines, respectively. Initial ($\Psi_i$) and final ($\Psi_f$) state's Faddeev amplitude for the involved baryons are represented by orange semi-circles. The blue blobs represent the corresponding quark-photon and diquark-photon vertices for each of the following diagrams: in $D_1$ it entails the photon coupling to a dressed-quark and in $D_2$ the photon couples elastically to a diquark.}
\end{figure}

The elastic electromagnetic form factors are obtained from the following microscopic baryon current:
\begin{align}
\mathcal{J}_{\mu,\rho\sigma}(P_f,P_i) &= \hspace{-.2cm} \sum_{n=1,2} \int_{dk} \mathcal{P}_{\rho \alpha}(P_f) \nonumber \\
&
\times \Lambda^{D_n}_{\mu,\alpha\beta} \left(k; P_f, P_i\right) \, \mathcal{P}_{\beta \sigma}(P_i) \,,
\end{align}
where $P_{i}$ and $P_f$ denote the incoming and outgoing momenta, respectively. Moreover, the integration notation entails $\int_{dk} = \int d^4k/(2\pi)^4$ and the sum is performed over the possible quark-photon and diquark-photon contributions $\Lambda^{D_n}_{\mu,\alpha\beta} \left(k; P_f, P_i\right)$ allowed in our quark-diquark picture, diagrammatically represented in Fig.~\ref{fig:EMInteractions}, and described in the subsequent subsections. Besides, Lorentz indices denote photon and $\Omega$ baryon polarizations: $\mu$ is associated with the photon whereas the remaining Greek letters are used for the $\Omega$ baryon. The positive energy projection operator is given by
\begin{equation}
\mathcal{P}_{\rho \alpha} (P) = \Lambda_{+}(P) \mathcal{R}_{\rho \alpha} (P) \,,
\end{equation}
where
\begin{align}
\Lambda_{+}(P) &= (1-i \gamma \cdot \hat{P})/2 \,, \\
\mathcal{R}_{\rho \alpha} (P) &= \delta_{\rho \alpha} -\frac{1}{3} \gamma_{\rho} \gamma_{\alpha} \nonumber \\
&
+\frac{2}{3} \hat{P}_{\rho} \hat{P}_{\alpha} +\frac{1}{3} \left( \gamma_{\rho} \hat{P}_{\alpha} - \gamma_{\alpha} \hat{P}_{\rho} \right) \,,
\end{align}
with $\hat{P} = P/m_\Omega$. 

Energy--momentum conservation, together with the on-shell conditions $P_i^2 = P_f^2 = -m^2_{\Omega}$, implies the fo\-llowing kinematic relations (with $Q = P_f -P_i$):
\begin{eqnarray}
Q \cdot P_f &=& +\frac{Q^2}{2} \,, \\
Q \cdot P_i &=& -\frac{Q^2}{2} \,, \\
P_f \cdot P_i &=& - m_\Omega^2 - \frac{Q^2}{2} \,.
\end{eqnarray}

Within the present framework, and retaining only axial-vector diquark correlations $\{ss\}_{1^+}$, current conservation is ensured by two classes of contributions: $D_i$-type ($i=1,2$) depicted in Fig.~\ref{fig:EMInteractions}. The microscopic baryon current can be written as
\begin{align}
\mathcal{J}_{\mu, \rho\sigma} \left( P_f, P_i \right) &= d^{\{ ss \}_{1^+}} \, e_{s} \, \mathcal{J}_{\mu,\rho\sigma}^{D_1} \left( P_f, P_i \right) \nonumber \\
&
+ d^{\{ss\}_{1^+}} \, e_{\left\{ ss \right\}} \, \mathcal{J}_{\mu,\rho\sigma}^{D_2} \left( P_f, P_i \right) \,, 
\label{microscopic current}
\end{align}
where $e_s=-1/3$ is the strange-quark charge (in units of the electron's charge) and $e_{\{ss\}}=-2/3$ the diquark charge.


\subsubsection{The diagram $D_1$: quark-photon coupling}

The first contribution corresponds to the photon coupling directly to a dressed strange quark inside the $\Omega$ baryon. The associated kernel reads
\begin{align}
\Lambda^{D_1}_{\mu,\alpha\beta} \left(k; P_f, P_i\right) &= e_s \, \mathcal{D}_{\alpha\delta}(P_f) \, S(-k_{-f}) \nonumber \\
&
\times \Lambda_{\mu}^{\gamma q} \, S(-k_{-i}) \, \Delta^{1^+}_{\delta\lambda}(-k) \, \mathcal{D}_{\lambda\beta}(P_i) \,,
\end{align}
where $k_{\pm i(f)} = -k \pm P_{i(f)}$. Moreover, the quark-photon vertex, $\Lambda_{\mu}^{\gamma q} \equiv \Lambda_{\mu}^{\gamma q}(Q)$, is defined consistently with the SCI and the RL approaches
\begin{eqnarray}
\Lambda_{\mu}^{\gamma q}(Q) = \gamma_{\mu}^{T} P_{T}(Q^2) + \eta \, \sigma_{\mu\nu}Q_{\nu} F_{AMM} \left( Q^2 \right) \,, \label{Quark-Photon vertex}
\end{eqnarray}
with $\gamma_{\mu}^{T} = \gamma_{\mu} - Q_{\mu} \gamma \cdot Q / Q^2$ and $\sigma_{\mu\nu}=i\left[ \gamma_{\mu}, \gamma_{\nu} \right]/2$, thus ensuring current conservation ($Q_{\mu} \mathcal{J}_{\mu,\rho\sigma} = 0$) for this contribution. The second term in Eq.~\eqref{Quark-Photon vertex}, modulated by a dimensionless factor $\eta$, encodes the effect of the large anomalous magnetic moment (AMM) characteristic of dressed quarks in the presence of DCSB:
\begin{eqnarray}
F_{AMM} \left( Q^2 \right) = \frac{1}{2M} \exp\left(-\frac{Q^2}{4M^2} \right) \,. \label{Vertex Dressing: quark-photon AMM} 
\end{eqnarray}
The transverse dressing function obtained from the symmetry-preserving SCI kernel is~\cite{Roberts:2011wy}
\begin{eqnarray}
\hspace{-.8cm} P_{T}^{-1}(Q^2) =&& \nonumber \\ && \hspace{-1.2cm} 1+\frac{4 \alpha_{IR}}{3 \pi m_g^2} \int_0^1{d\alpha \, \alpha \left( 1- \alpha \right) Q^2 \mathcal{C}_2 \left( \omega \left( \alpha, Q^2 \right) \right) } , \label{PT from SCI}
\end{eqnarray}
with $\omega \left( \alpha ,Q^2 \right) = M^2 + \alpha \left( 1- \alpha \right) Q^2$ and
\begin{eqnarray}
\mathcal{C}_{\beta} (x) =  \frac{\omega ^{2-\beta}}{\Gamma \left[ \beta \right]} \Gamma \left[ \beta -2, x \, \tau_{uv}^2, x \, \tau_{ir}^2 \right] \,,
\end{eqnarray}
where the gamma and generalized incomplete gamma functions are used.

\begin{figure}[!t]
\includegraphics[clip, trim={0.0cm 0.0cm 0.0cm 0.0cm}, width=0.45\textwidth]{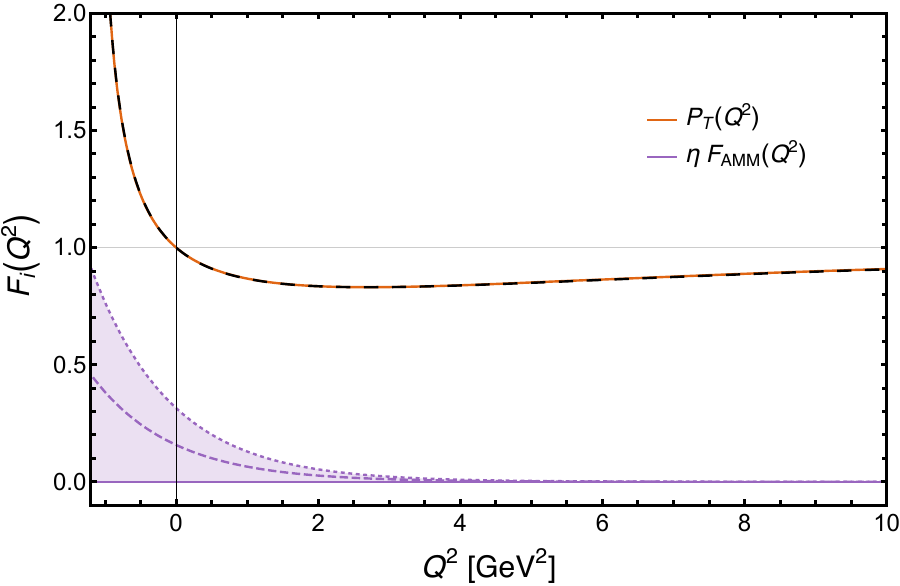}
\caption{\label{fig:PTcomparison} Transverse quark--photon dressing functions appearing in Eq.~\eqref{Quark-Photon vertex}. The rust red solid line is the SCI result, Eq.~\eqref{PT from SCI}, and the black long-dashed line is the parametrization shown in Eq.~\eqref{PT Fit}. At large space-like $Q^2$, the pointwise behavior yields $P_T(Q^2)\rightarrow 1$, characteristic of a pointlike particle. The $Q^2$ dependence of $F_{AMM}$ in Eq.~\eqref{Vertex Dressing: quark-photon AMM} is also shown for three representative values of the dressed-quark anomalous magnetic moment parameter, $\eta = 0$ (solid line), $\eta = 1/6$ (dashed line), and $\eta = 1/3$ (dotted line).}
\end{figure}

An amenable approximation for $P_T(Q^2)$ that parametrizes quite well the solution of Eq.~\eqref{PT from SCI} on the region $Q^2 / \hbox{GeV}^2 \in \left[ -1.26 , 10 \right]$ is shown in Fig.~\ref{fig:PTcomparison}. This fit entails a pole at $Q^2=-1.262 \, \hbox{GeV}^2$, consonant with the $\phi$-pole, $m_{\phi}^{\rm SCI}=1.13\,\hbox{GeV}$~\cite{Yin:2021uom},  and it is parametrized as
\begin{eqnarray}
P_T \left( x \right) = \frac{1+ 0.6473 \, x + 0.1000 \, x^2 + 0.0111 \, x^3}{1 + 0.9103 \, x + 0.1069 \, x^2 + 0.0107 \, x^3} \,. \label{PT Fit}
\end{eqnarray}
Note also that Fig.~\ref{fig:PTcomparison} shows the $Q^2$-dependence of $F_{AMM}$ depicted in Eq.~\eqref{Vertex Dressing: quark-photon AMM} for three representative values of the dressed-quark anomalous magnetic moment parameter, $\eta = 0$ (solid line), $\eta = 1/6$ (dashed line), and $\eta = 1/3$ (dotted line).


\begin{table*}[!t]
\centering
\begin{tabular}[t]{c|r r r r|r r r r|r}
\hline\hline
\tstrut
& \; $a_0$ \; & \; $a_1$ \; & \; $b_1$ \; & \; $b_2$ \; & \; $a'_0$ \; & \; $a'_1$ \; & \; $b'_1$ \; & \; $b'_2$ \; & \;  $\upsilon$ \; \\
\hline
\tstrut
\; $F_1$ \; & \; 1.000 \; & \; 0.317 \; & \; 1.236 \; & \; 0.359 \; & \; 0.000 \; & \; -0.592 \; & \; 0.581 \; & \; 0.0255 \; & \; 0.868 \; \\
\; $F_2$ \; & \; -2.026 \; & \; -0.519 \; & \; 1.101 \; & \; 0.238 \; & \; -2.075 \; & \; -0.668 \; & \; 0.559 \; & \; 0.035 \; & \; 1.349 \; \\
\; $F_3$ \; & \; 0.312 \; & \; 0.021 \; & \; 1.089 \; & \; 0.248 \; & \; 1.714 \; & \; -1.157 \; & \; -0.241 \; & \; -0.294 \; & \; 0.931 \; \\
\hline\hline
\end{tabular}
\caption{\label{Vertex Dressing Function Parameters} Interpolation coefficients for each respective elastic diquark--photon form factor.}
\end{table*}

\begin{figure}[!t]
\includegraphics[clip, trim={0.0cm 0.0cm 0.0cm 0.0cm}, width=.48\textwidth]{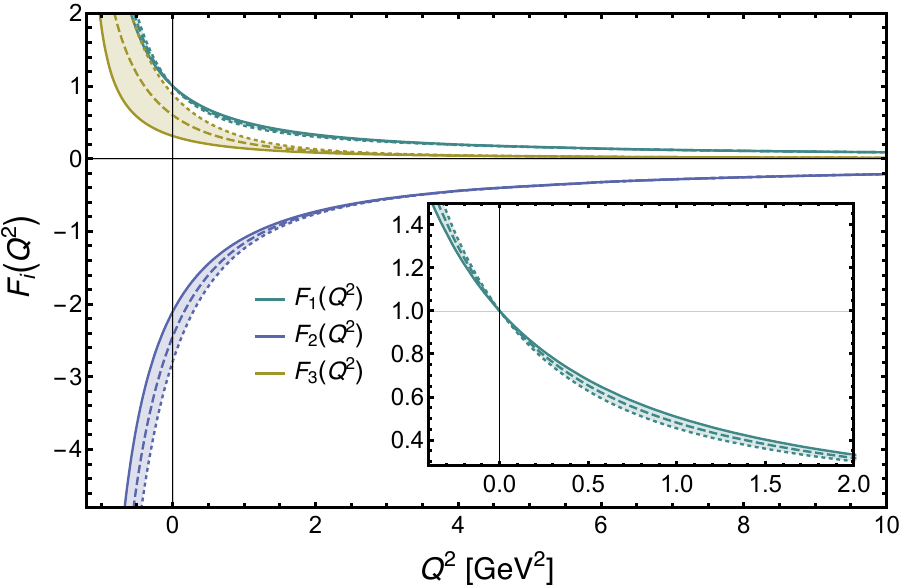}
\caption{\label{fig:Diquark-Photon} Axial-vector diquark-photon dressing functions. Our results are depicted for three representative values of the dressed-quark anomalous magnetic moment parameter, $\eta = 0$ (solid line), $\eta = 1/6$ (dashed line), and $\eta = 1/3$ (dotted line).}
\end{figure}

\subsubsection{The diagram $D_2$: elastic diquark-photon coupling}

The second contribution corresponds to the elastic photon scattering from the axial-vector diquark inside the $\Omega$ baryon. The associated kernel is
\begin{align}
\Lambda^{D_2}_{\mu,\alpha\beta} \left(k; P_f, P_i\right) &=
e_{\{ss\}} \, \mathcal{D}_{\alpha\delta}(P_f) \, S(k) \, \Delta^{1^+}_{\delta\rho}(k_f) \nonumber \\
&
\times \Lambda_{\mu,\rho\sigma}^{\gamma dq} \, \Delta^{1^+}_{\sigma\lambda}(k_i) \, \mathcal{D}_{\lambda\beta}(P_i) \,,
\end{align}
where the structure of the corresponding diquark-photon vertex, $\Lambda_{\mu,\rho\sigma}^{\gamma dq} \equiv \Lambda_{\mu,\rho\sigma}^{\gamma dq}(k_f, k_i)$, is 
\begin{eqnarray}
i \Lambda_{\mu, \rho \sigma}^{\gamma dq}(k_f, k_i) = \sum_{j=1}^{3} {T^j_{\mu, \rho \sigma} \left( K,Q \right) F_j(Q^2)} \,, 
\label{Vertex: photon-ax diquark}
\end{eqnarray}
with
\begin{eqnarray}
T^1_{\mu, \rho \sigma} \left( K,Q \right) &=& 2 K_{\mu} \mathcal{T}^{\,i}_{\rho \nu} \mathcal{T}^f_{\nu \sigma} \,, \\
T^2_{\mu, \rho \sigma} \left( K,Q \right) &=& Q_{\nu} \mathcal{T}_{\rho \nu} \left( k_i, -Q/2 \right) \mathcal{T}^f_{\mu \sigma} \nonumber \\
&& 
\hspace*{-0.50cm} - Q_{\nu} \mathcal{T}_{\sigma \nu} \left( k_f, Q/2 \right) \mathcal{T}^i_{\mu \rho} \,, \\
T^3_{\mu, \rho \sigma} \left( K,Q \right) &=& \frac{K_{\mu} Q_{\nu} Q_{\lambda}}{m_{1^{+}}^2} \nonumber \\
&&
\hspace*{-0.50cm} \times \mathcal{T}_{\rho \nu} \left( k_i, -Q/2 \right) \mathcal{T}_{\lambda \nu} \left( k_f, Q/2 \right) \,,
\end{eqnarray}
where $\mathcal{T}_{\rho \sigma} \left( k, p \right) = \delta_{\rho \sigma} + k_{\rho} p_{\sigma} / m_{1^{+}}^2$ and $\mathcal{T}^{i(f)}_{\rho \sigma}  = \mathcal{T}_{\rho \sigma} \left( k_{i(f)}, k_{i(f)} \right)$. 

The dressing functions in Eq.~\eqref{Vertex: photon-ax diquark} can be split into a $\phi$-pole (PP) and an AMM contribution,  
\begin{eqnarray}
F_{j}(Q^2) = F_{PP} \hspace{-.1cm} \left( Q^2 \right) + \eta \, \tilde{F}_{AMM} \hspace{-.1cm} \left( Q^2 \right) \,, \label{Vertex Dressings decomposition}
\end{eqnarray}
such that
\begin{eqnarray}
\hspace{-.3cm} F_{PP} \left( x \right) \hspace{-.1cm} &=& \hspace{-.1cm} \frac{a_0 + a_1 x }{1 +b_1 x + b_2 x^2} \,, \label{Vertex Dressing: RP-contribution} \\
\hspace{-.3cm} \tilde{F}_{AMM} \left( x \right) \hspace{-.1cm} &=& \hspace{-.1cm} \frac{a'_0 + a'_1 x}{1 +b'_1 x + b'_2 x^2}  \exp \left( -\upsilon x \right) \,, \label{Vertex Dressing: AMM-contribution}
\end{eqnarray}
with the associated fit parameters listed in Table~\ref{Vertex Dressing Function Parameters}. These were already guessed while studying the transition electromagnetic form factors of the nucleon's first excited and parity partner baryons: $N(1440)\frac{1}{2}^+$~\cite{Wilson:2011aa}, $\Delta(1232)\frac{3}{2}^+$~\cite{Segovia:2013uga}, $N(1535)\frac{1}{2}^-$~\cite{Raya:2021pyr}, $\Delta(1700)\frac{3}{2}^-$~\cite{Albino:2025fcp} and $N(1520)\frac{3}{2}^-$~\cite{Albino:2025bnr}.

We note that, although the parameter $\eta$ originates from the quark--photon interaction~\cite{Xing:2021dwe}, its impact may differ between quark- and diquark-level contributions. In this work, we consider the same value of the dressed-quark anomalous magnetic moment parameter in both the quark-photon and diquark-photon vertices. The resulting $\eta$-dependence of the axial-vector diquark-photon dressing functions is illustrated in Fig.~\ref{fig:Diquark-Photon} for three representative values: $\eta = 0$ (solid line), $\eta = 1/6$ (dashed line), and $\eta = 1/3$ (dotted line).


\subsection{Elastic electromagnetic current}
\label{subsec:Current}

For a spin-3/2 baryon such as the $\Omega$, the electromagnetic current can be decomposed in terms of four independent form factors. In our notation, the current reads
\begin{equation}
\mathcal{J}_{\mu, \rho \sigma} \left( P_f, P_i \right) =  i\, \mathcal{P}_{\rho \alpha}  \hspace{-.1cm} \left( P_f \right) \Gamma_{\mu, \alpha \beta} \left( P_f, P_i \right) \mathcal{P}_{\beta \sigma} \hspace{-.1cm} \left( P_i \right),
\end{equation}
where the vertex function $\Gamma_{\mu,\alpha\beta}$ is given by
\begin{align}
&
\Gamma_{\mu, \alpha \beta} \left( P_f, P_i \right) = \left[ F^{*}_1 \left( Q^2 \right) \gamma_{\mu} + \frac{1}{2 m_\Omega} F^{*}_2 \left( Q^2 \right) \sigma_{\mu \nu} Q_{\nu} \right] \delta_{\alpha \beta} \nonumber \\
& + 2\left[ F^{*}_3 \left( Q^2 \right) \gamma_{\mu} +\frac{1}{2 m_\Omega} F_4^{*} \left( Q^2 \right) \sigma_{\mu \nu} Q_{\nu} \right] \frac{Q_{\alpha} Q_{\beta}}{4 m_{\Omega}^2} \,.
\label{Delta-Photon Vertex} 
\end{align}

The form factors $F_i^\ast(Q^2)$, with $i=1,\ldots,4$, provide a complete description of the electromagnetic structure of the $\Omega$ baryon. However, it is often convenient to express the current in terms of multipole form factors, which admit a more direct physical interpretation. These are the electric charge $G_{E0}$, magnetic dipole $G_{M1}$, electric quadrupole $G_{E2}$, and magnetic octupole, $G_{M3}$. These multipoles are related to the $F_i^\ast$ through the linear combinations
\begin{align}
G_{E0} \left( Q^2 \right) &= \left( 1+ \frac{2 \tau_{\Omega}}{3} \right) \Delta F^{*}_{1,2} \nonumber \\
&
- \frac{\tau_{\Omega}}{3} \left( 1+ \tau_{\Omega} \right) \Delta F^{*}_{3,4} \,, \label{GE0 Omega} \\
G_{M1} \left( Q^2 \right) &= \left( 1+ \frac{4 \tau_{\Omega}}{5} \right) \Sigma F^{*}_{1,2} \nonumber \\
&
- \frac{2 \tau_{\Omega}}{5} \left( 1+ \tau_{\Omega} \right) \Sigma F^{*}_{3,4} \,, \label{GM1 Omega} \\
G_{E2} \left( Q^2 \right) &= \Delta F^{*}_{1,2}  - \frac{1}{2} \left( 1+ \tau_{\Omega} \right) \Delta F^{*}_{3,4} \,, \label{GE2 Omega} \\
G_{M3} \left( Q^2 \right) &= \Sigma F^{*}_{1,2} - \frac{1}{2} \left( 1+ \tau_{\Omega} \right) \Sigma F^{*}_{3,4} \,, \label{GM3 Omega}
\end{align}
where we have introduced the shorthand notation
\begin{equation}
\Sigma F^{*}_{i,j} = F^{*}_i + F^{*}_j \,, \quad\quad \Delta F^{*}_{i,j} = F^{*}_i - \tau_{\Omega} F^{*}_j \,,
\end{equation} 
with $\tau_{\Omega} = Q^2/(4m_{\Omega}^2)$. 

Current conservation requires that the electric charge form factor satisfies
\begin{equation}
G_{E0}(0)=-1 \,,
\end{equation}
which fixes the overall normalization of the $\Omega$ baryon current, and yields a charge radius 
\begin{eqnarray}
r_\Omega \equiv \sqrt{-6 G'_{E0}(0)/G_{E0}(0)} = (0.555 \pm_{\eta} 0.026) \, \hbox{fm} \,,
\end{eqnarray}
consistent with lattice-QCD determinations reported in Refs.~\cite{Boinepalli:2009sq,Alexandrou:2010jv}: $(0.554 \pm 0.014) \, \hbox{fm}$ and $(0.590 \pm 0.044) \, \hbox{fm}$, respectively. Moreover, the values of the multipole form factors at $Q^2=0$ define the static electromagnetic moments of the $\Omega$ baryon. In particular, the magnetic dipole moment, electric quadrupole moment, and magnetic octupole moment are respectively given by
\begin{align}
\hat{\mu}_{\Omega} &\equiv G_{M1}(0) \cdot \frac{m_N \mu_N}{m_\Omega} = (-2.742 \pm_{\eta} 0.359)\,\mu_N \,, \label{magnetic dipole moment} \\ 
\hat{\mathcal{Q}}_{\Omega} &\equiv G_{E2}(0) \cdot \frac{|e|}{m_{\Omega}^2} = (+0.0211 \pm_{\eta} 0.0039)\,\text{fm}^2 \,, \label{electric quadrupole moment} \\ 
\hat{\mathcal{O}}_{\Omega} &\equiv G_{M3}(0) \cdot \frac{m_N^3 {\cal O}_N}{ m_\Omega^3} = (-0.344 \pm_{\eta} 0.052)\,{\cal O}_N \,. \label{magnetic octupole moment} 
\end{align}
The measured magnetic moment, $\mu_{\Omega}^{\rm e}=(-2.02\pm0.05)\,\mu_N$~\cite{ParticleDataGroup:2024cfk}, is the best determined electromagnetic observable of the $\Omega^-$. The central result in Eq.~\eqref{magnetic dipole moment} is about $36\%$ larger in magnitude and its variation band does not overlap the experimental interval. Moreover, increasing the positive AMM parameter increases $|\mu_\Omega|$ and hence moves the result further away from experiment. Thus, within the present truncation, the empirical moment favors a small AMM contribution and points to missing dynamics in the bound-state amplitude and/or the quark--photon vertex. The remaining moments may be compared with other theoretical determinations in Fig.~6 of Ref.~\cite{Fu:2025vkq}.

All static observables are quoted at the central value $\eta=1/6$. The displayed spreads are generated solely by varying $\eta\in[0,1/3]$ and quantify sensitivity to this phenomenological parameter; they are not statistical or systematic error estimates and, for that reason, they are denoted as ``$\pm_\eta$". In particular, they do not include uncertainties associated with the contact interaction, the static approximation, the quark--diquark reduction, the omitted meson cloud, or the restricted vertex Ansatz.


\subsection{Time-like extrapolation}
\label{subsec:TimeLikeFFs}

To explore the continuation of the $\Omega$ form factors into the time-like region, we follow Ref.~\cite{Fu:2025vkq} and relate them to their space-like counterparts using an asymptotically motivated prescription~\cite{Ramalho:2020laj}. In particular, the extrapolated multipole form factors are defined by
\begin{align}
G_{E(0,2)}^{\text{TL}}(q^2) &= G_{E(0,2)}(Q^2 + 2 m_{\Omega}^2) \,, \label{SL_to_TL01} \\
G_{M(1,3)}^{\text{TL}}(q^2) &= G_{M(1,3)}(Q^2 + 2 m_{\Omega}^2) \,, \label{SL_to_TL02}
\end{align}
with $q^2 = -Q^2 > 0$. The shift of $2m_\Omega^2$ in the arguments reflects the difference between the space-like normalization point ($Q^2=0$) and the physical time-like threshold ($q^2=4m_\Omega^2$).

These relations are motivated by general properties of the electromagnetic current, including unitarity and analyticity, together with the Phragm\'en--Lindel\"of theorem, which implies that the asymptotic behavior of form factors is identical in the space-like and time-like regions~\cite{Denig:2012by, Pacetti:2014jai}.

The resulting functions are real and do not contain the complex phases generated by intermediate hadronic states, final-state interactions, resonances, or unitarity cuts. Hence, this prescription is not a dynamical calculation of the physical time-like analytic structure and should not be applied quantitatively near threshold.

The relations are justified only in the asymptotic regime of large $|q^2|$; their use at moderate energies is therefore a model-dependent extrapolation. We employ it to construct the effective form factor and compare its broad energy dependence with data, while keeping these limitations explicit.


\section{RESULTS}
\label{sec:results}

\begin{figure}[!t]
\includegraphics[clip, trim={0.0cm 0.0cm 0.0cm 0.0cm}, width=0.48\textwidth]{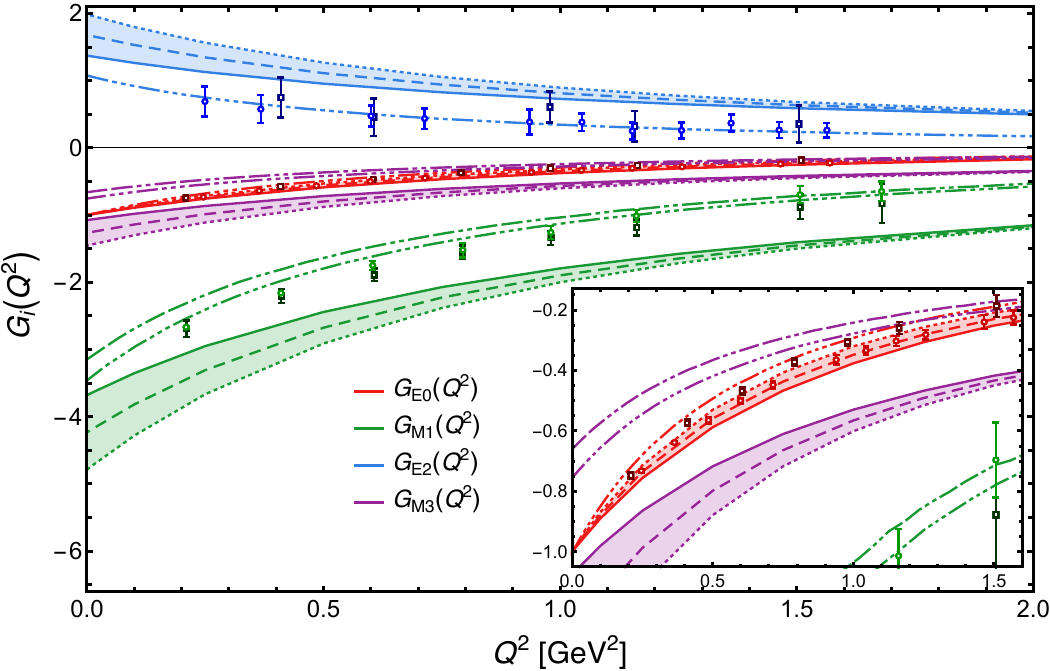}
\caption{\label{fig:Space-likeFFs} Space-like electromagnetic form factors of the $\Omega$ baryon. Red band: electric charge form factor $G_{E0}(Q^2)$. Green band: magnetic dipole form factor $G_{M1}(Q^2)$. Blue band: electric quadrupole form factor $G_{E2}(Q^2)$. Purple band: magnetic octupole form factor $G_{M3}(Q^2)$. Results are shown for three representative values of the dressed-quark anomalous magnetic moment parameter, $\eta = 0$ (solid line), $\eta = 1/6$ (dashed line), and $\eta = 1/3$ (dotted line). Double-dot-dashed and triple-dot-dashed curves of the corresponding color depict the calculations of Ref.~\cite{Fu:2025vkq} obtained without and with kaon-cloud contributions, respectively: kaon-cloud effects are found to be negligible for $G_{E0}(Q^2)$ and $G_{E2}(Q^2)$, so only a single curve is visible in these cases. Data points are the latest lattice QCD calculations of the electromagnetic form factors of the $\Omega$ baryon~\cite{Alexandrou:2010jv}.
}
\end{figure}

\subsection{Form factors in the space-like region}
\label{subsec:space-like}

We now present our results for the elastic electromagnetic form factors of the $\Omega$ baryon in the space-like region, $Q^2 \geq 0$. The four independent multipole form factors, namely the electric monopole $G_{E0}$, magnetic dipole $G_{M1}$, electric quadrupole $G_{E2}$, and magnetic octupole $G_{M3}$, are displayed in Fig.~\ref{fig:Space-likeFFs}. The bands quantify the sensitivity to the dressed-quark anomalous magnetic moment, obtained by varying the parameter $\eta \in [0,1/3]$, with the central curve corresponding to $\eta = 1/6$.

By construction, the electric monopole form factor satisfies the normalization condition $G_{E0}(0) = -1$, thereby ensuring electromagnetic current conservation. As the momentum transfer increases, $G_{E0}(Q^2)$ evolves smoothly toward zero, reflecting the finite spatial extent of the $\Omega$ baryon. The relatively slow falloff observed in Fig.~\ref{fig:Space-likeFFs} is characteristic of contact-interaction frameworks, where the absence of explicit momentum dependence in the interaction kernel typically produces harder form factors than those obtained with QCD-based running interactions. Nevertheless, the overall behavior remains consistent with expectations for a compact three-quark bound state dominated by strange degrees of freedom.

The magnetic dipole form factor $G_{M1}(Q^2)$ remains negative throughout the kinematic domain explored here, as expected for a negatively charged baryon composed solely of strange quarks. Its value at $Q^2=0$ determines the magnetic dipole moment quoted in Eq.~\eqref{magnetic dipole moment}. In contrast with $G_{E0}$, the magnetic form factor exhibits a visible sensitivity to the dressed-quark anomalous magnetic moment. Increasing $\eta$ enhances the magnitude of $G_{M1}$, particularly at low and intermediate momentum transfers, indicating that dynamical chiral symmetry breaking effects encoded in the quark AMM play an important role in shaping the magnetic structure of the $\Omega$ baryon.

The electric quadrupole form factor $G_{E2}(Q^2)$ is found to be non-vanishing and positive over the full kinematic domain shown in Fig.~\ref{fig:Space-likeFFs}. Within the adopted convention, this behavior signals the presence of non-spherical components in the charge distribution of the spin-$3/2$ system. The quadrupole form factor exhibits a moderate dependence on the parameter $\eta$, with the dominant effect being an overall enhancement of its magnitude rather than a change in its qualitative behavior. This is consistent with the expectation that higher multipole moments probe more detailed aspects of the internal spin and orbital structure.

The magnetic octupole form factor $G_{M3}(Q^2)$ is also nonzero and remains negative in the entire range of momentum transfer considered. As shown in Fig.~\ref{fig:Space-likeFFs}, the dependence of $G_{M3}$ on the dressed-quark AMM is more pronounced at low $Q^2$, where variations in $\eta$ produce visible modifications in its magnitude. Such sensitivity is not unexpected, since higher-order magnetic multipoles are naturally more sensitive to subleading dynamical mechanisms and to the details of the electromagnetic current construction.

Overall, the inclusion of the dressed-quark anomalous magnetic moment produces a systematic enhancement of the magnetic and higher-order multipole form factors, while leaving the qualitative behavior of the electric monopole form factor largely unchanged. This highlights the non-negligible role of dynamical chiral symmetry breaking effects in the electromagnetic structure of the $\Omega$ baryon, even in a system composed solely of strange quarks.

It is worth emphasizing that the general trends displayed in Fig.~\ref{fig:Space-likeFFs} are qualitatively consistent with previous continuum studies of decuplet baryons and with available lattice-QCD simulations. Quantitative differences are nevertheless expected. In particular, the contact interaction employed herein lacks the momentum dependence associated with QCD's running interaction and therefore typically generates form factors that are harder than those obtained using momentum-dependent kernels. Furthermore, the present calculation describes the dressed-quark core of the $\Omega$ baryon and does not incorporate explicit meson-cloud effects, which are known to influence the low-$Q^2$ behavior of electromagnetic observables. Consequently, these aspects should be kept in mind when comparing our results with lattice-QCD data, as in Fig.~\ref{fig:Space-likeFFs}, and future experimental measurements.


\begin{figure}[!t]
\includegraphics[clip, trim={0.0cm 0.0cm 0.0cm 0.0cm}, width=0.45\textwidth]{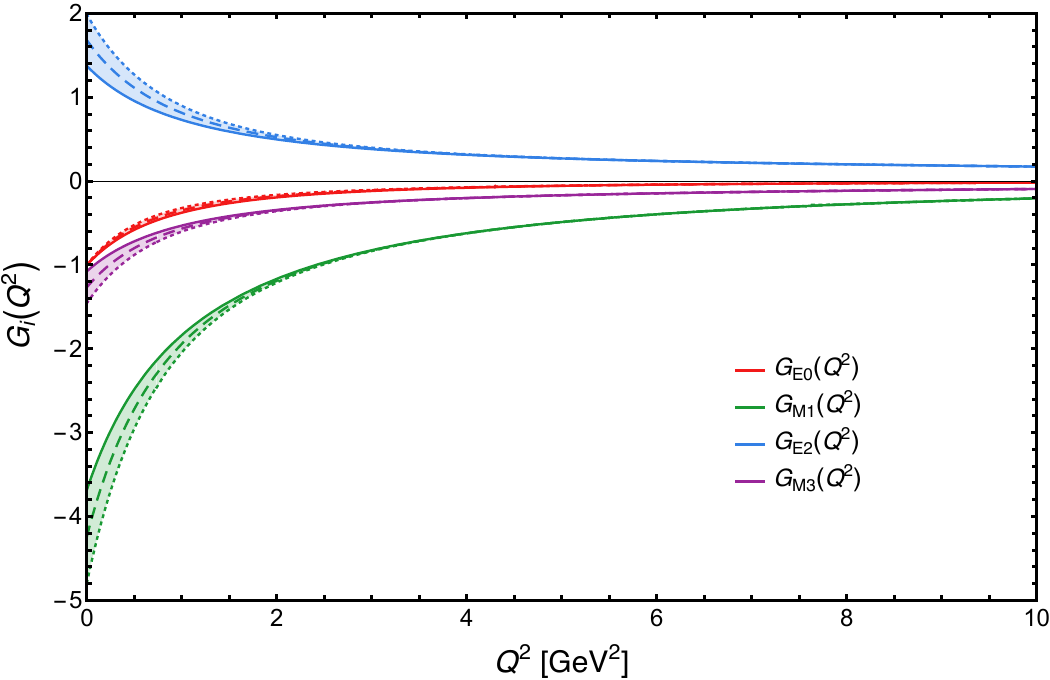}
\caption{\label{AsymptoticSpace-likeFFs} Space-like electromagnetic form factors of the $\Omega$ baryon over the extended kinematic domain $Q^2 \in [0,10]~{\rm GeV}^2$. The curves correspond to the fit forms given in Eqs.~(\ref{eq:fit1}-\ref{eq:fit3}). The bands are generated by varying the dressed-quark anomalous magnetic moment parameter within $\eta \in [0,1/3]$, with the central curve corresponding to $\eta = 1/6$. The figure illustrates the asymptotic falloff of the multipole form factors, with $G_{E0}$ and $G_{M1}$ behaving approximately as $1/Q^4$, whereas $G_{E2}$ and $G_{M3}$ exhibit a stronger suppression consistent with a $1/Q^6$ behavior.}
\end{figure}

\begin{figure}[!t]
\includegraphics[clip, trim={0.0cm 0.0cm 0.0cm 0.0cm}, width=0.45\textwidth]{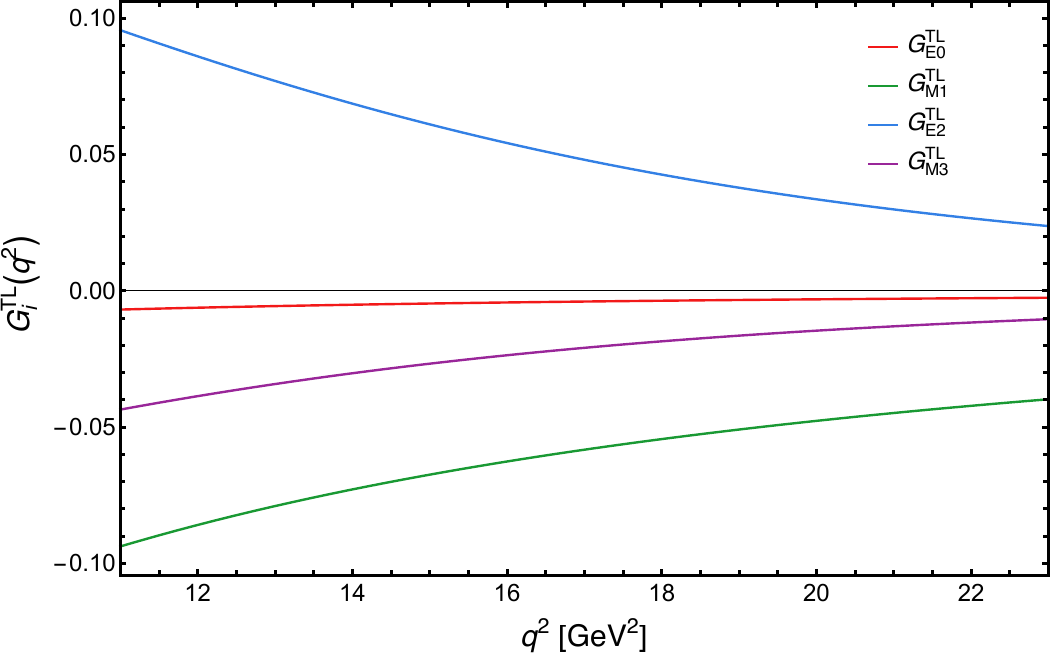}
\caption{\label{Time-likeFFs} Time-like electromagnetic form factors of the $\Omega$ baryon obtained through the analytic continuation described in Sec.~\ref{subsec:TimeLikeFFs}. Red band: electric monopole form factor $G^{\rm TL}_{E0}(q^2)$. Green band: magnetic dipole form factor $G^{\rm TL}_{M1}(q^2)$. Blue band: electric quadrupole form factor $G^{\rm TL}_{E2}(q^2)$. Purple band: magnetic octupole form factor $G^{\rm TL}_{M3}(q^2)$. The curves correspond to $\eta = 1/6$; for variations within $\eta \in [0,1/3]$, the effect of the dressed-quark anomalous magnetic moment is negligible in the displayed time-like region and therefore does not produce visible bands.}
\end{figure}

\subsection{Extrapolation into the time-like region}
\label{subsec:time-like}

We now consider the model-dependent extrapolation of the $\Omega$ form factors into the time-like region. Since the prescription introduced in Sec.~\ref{subsec:TimeLikeFFs} is motivated by asymptotic properties, it is first useful to examine
the large-$Q^2$ behavior of the corresponding space-like form factors.

The space-like multipole form factors  over the extended kinematic domain $Q^2 \in [0,10]~{\rm GeV}^2$ are displayed in Fig.~\ref{AsymptoticSpace-likeFFs}. The asymptotic behavior of the form factors can be described by simple algebraic parametrizations. In particular, our numerical results indicate that the leading multipoles approximately follow
\begin{equation}
G_{E0}(Q^2),\, G_{M1}(Q^2) \propto \frac{1}{Q^4} \,,
\label{eq:asymp1}
\end{equation}
whereas the higher multipoles behave approximately as
\begin{equation}
G_{E2}(Q^2),\, G_{M3}(Q^2) \propto \frac{1}{Q^6} \,.
\label{eq:asymp2}
\end{equation}
Although the present contact-interaction framework does not reproduce the logarithmic scaling violations expected in perturbative QCD, the observed power-law suppression is qualitatively consistent with general counting-rule expectations for baryonic form factors and exhibits the same large-$Q^2$ behavior reported in Ref.~\cite{Ramalho:2020laj}.

In order to provide compact representations of the numerical solutions, we employ the parametrizations
\begin{equation}
G_{i} \hspace{-.1cm} \left( Q^2 \right) = G_{i,PP} \hspace{-.1cm} \left( Q^2 \right) + \eta \, \tilde{G}_{i,AMM} \hspace{-.1cm} \left( Q^2 \right) \,,
\label{eq:fit1}
\end{equation}
with
\begin{align}
\hspace{-.3cm} G_{i,PP} \left( x \right) \hspace{-.1cm} &= \hspace{-.1cm} \frac{a_0 + a_1 x}{1 +b_1 x + b_2 x^2 + b_3 x^3 + b_4 x^4} \,, \label{eq:fit2} \\
\hspace{-.3cm} \tilde{G}_{i,AMM} \left( x \right) \hspace{-.1cm} &= \hspace{-.1cm} \frac{a'_0 + a'_1 x + a'_2 x^2}{1 + b'_1 x + b'_2 x^2 + b'_3 x^3}  \exp \left( -\upsilon x \right) \,.
\label{eq:fit3}
\end{align}
The corresponding fit parameters are listed in Table~\ref{tab:AsyCoe-SLFFs}. These expressions provide an accurate interpolation of the numerical results across the entire kinematic domain considered in this work while simultaneously reproducing the observed asymptotic behavior.

\begin{table*}[!t]
\centering
\scalebox{0.85}{
\begin{tabular}[t]{c|rrrrrr|rrrrrr|r}
\hline\hline
\tstrut
& \; $a_0$ \; & \; $a_1$ \; & \; $b_1$ \; & \; $b_2$ \; & \; $b_3$ \; & \; $b_4$ \; & \; $a'_0$ \; & \; $a'_1$ \; & \; $a'_2$ \; & \; $b'_1$ \; & \; $b'_2$ \; & \; $b'_3$ \; & \;  $\upsilon$ \; \\
\hline
\tstrut
\; $G_{E0}$ \; & \; -1 \; & \; 0 \; & \; 1.2388 \; & \; 0.4174 \; & \; 0 \; & \; 0 \; & \; 0 \; & \; 0.7537 \; & \; -0.1757 \; & \; 0.5843 \; & \; -0.0010 \; & \; -0.0419 \; & \; 0.8046 \; \\
\; $G_{M1}$ \; & \; -3.6787 \; & \; 0 \; & \; 0.9284 \; & \; 0.0757 \; & \; 0 \; & \; 0 \; & \; -3.3233 \; & \; 0.9373 \; & \; -0.0565 \; & \; 0.7803 \; & \; 0.1480 \; & \; -0.0087 \; & \; 0.7163 \; \\
\; $G_{E2}$ \; & \; 1.372 \; & \; 0.1558 \; & \; 0.9725 \; & \; 0.1362 \; & \; -0.0134 \; & \; 0.0006 \; & \; 1.8417 \; & \; -0.4205 \; & \; 0.0268 \; & \; 0.7842 \; & \; 0.1604 \; & \; 0 \; & \; 0.3911 \; \\
\; $G_{M3}$ \; & \; -1.0749 \; & \; -0.1151 \; & \; 1.0600 \; & \; 0.2005 \; & \; -0.0168 \; & \; 0.0009 \; & \; -1.1388 \; & \; 1.5391 \; & \; 0 \; & \; -0.9468 \; & \; -0.5760 \; & \; 0.0394 \; & \; 1.3552 \; \\
\hline\hline
\end{tabular}
}
\caption{\label{tab:AsyCoe-SLFFs} Interpolation coefficients for each space-like form factor at large transferred momenta.}
\end{table*}

The extrapolated time-like form factors are shown in Fig.~\ref{Time-likeFFs}. By construction, Eqs.~\eqref{SL_to_TL01} and~\eqref{SL_to_TL02} preserve the hierarchy of the space-like multipoles and produce functions that evolve smoothly above threshold. This smoothness is a property of the extrapolation and should not be interpreted as a prediction that physical threshold or resonance structures are absent.

The electric monopole form factor remains negative over the explored kinematic region, while the electric quadrupole form factor remains positive. Similarly, both magnetic multipoles preserve their negative sign after analytic continuation. Their overall magnitudes decrease progressively with increasing $q^2$, consistently with the asymptotic constraints discussed above.

The dependence on the dressed-quark anomalous magnetic moment becomes comparatively weaker in the time-like region. This behavior is expected because the AMM contribution primarily modifies the low-$Q^2$ structure of the electromagnetic current and therefore has a reduced impact at the comparatively large momenta relevant to the present time-like analysis.

As explained in Sec.~\ref{subsec:TimeLikeFFs}, these real functions do not include physical phases, hadronic singularities, or unitarity effects. Figure~\ref{Time-likeFFs} therefore displays an asymptotically motivated extrapolation, not a complete prediction for the physical time-like multipoles.


\begin{figure}[!t]
\includegraphics[clip, trim={0.0cm 0.0cm 0.0cm 0.0cm}, width=0.45\textwidth]{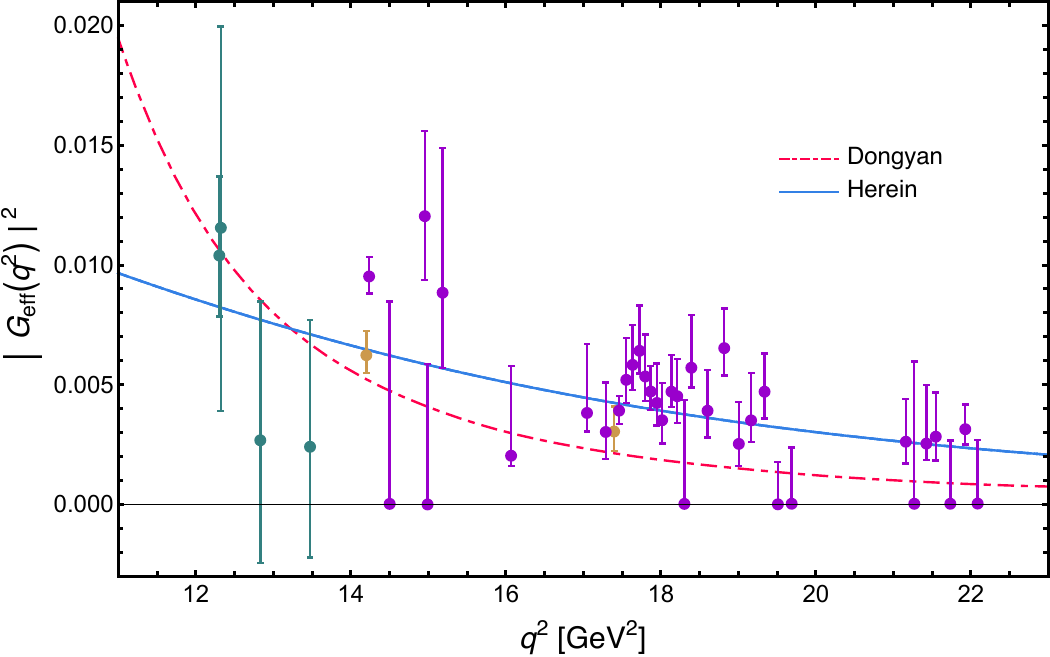}
\caption{\label{fig:EffectiveFFs} Effective form factor of the $\Omega$ baryon as a function of time-like $q^2$. The solid curve is obtained from the model-dependent extrapolation of the space-like form factors, while the dash-dotted curve corresponds to the parametrization of Ref.~\cite{Fu:2025vkq}, which includes meson-cloud effects. Experimental data points are also shown for comparison: yellow~\cite{Dobbs:2017hyd}, green~\cite{BESIII:2022kzc}, and purple~\cite{BESIII:2025azv}.}
\end{figure}

\subsection{Effective form factor and comparison with experiment}
\label{subsec:Effective}

In the time-like region, the individual multipole form factors cannot yet be disentangled experimentally. Consequently, present information on the electromagnetic structure of the $\Omega$ baryon is commonly encoded in terms of an effective form factor extracted from the total cross section for the reaction $e^+e^- \to \Omega^- \bar{\Omega}^+$.

For a spin-$3/2$ baryon, such as the $\Omega$, the effective form factor can be written as~\cite{Ramalho:2020laj, Fu:2025vkq}
\begin{align}
\left|G_{\rm eff}(q^2)\right|^2
&=
\left(1+\frac{1}{2\tau}\right)^{-1} \nonumber \\
&
\hspace*{-0.5cm} \times \Bigg[ \frac{10}{9} \left|G^{\rm TL}_{M1}(q^2)\right|^2 + \frac{16}{15} \tau^2 \left|G^{\rm TL}_{M3}(q^2)\right|^2 \nonumber\\
&
\hspace*{-0.5cm} +\frac{1}{2\tau} \left(2\left|G^{\rm TL}_{E0}(q^2)\right|^2 + \frac{8}{9} \tau^2 \left|G^{\rm TL}_{E2}(q^2)\right|^2 \right) \Bigg] \,,
\label{eq:Geff}
\end{align}
where
\begin{equation}
\tau = \frac{q^2}{4m_\Omega^2}\, .
\end{equation}
This quantity constitutes the observable directly accessible through the integrated cross section and therefore provides the most direct connection between theoretical calculations and the available experimental data.

The effective form factor obtained within the present framework is displayed in Fig.~\ref{fig:EffectiveFFs}, together with the phenomenological parametrization of Ref.~\cite{Fu:2025vkq} and the available experimental measurements~\cite{Dobbs:2017hyd, BESIII:2022kzc, BESIII:2025azv}. Within our calculation, the effective form factor is dominated by the electric monopole and magnetic dipole contributions, whereas the electric quadrupole and magnetic octupole provide comparatively small corrections.

As shown in Fig.~\ref{fig:EffectiveFFs}, the extrapolated effective form factor follows the broad decreasing trend of the data at intermediate and larger values of $q^2$. Given the model dependence of Eqs.~\eqref{SL_to_TL01} and~\eqref{SL_to_TL02}, this comparison should be regarded as qualitative and does not establish that all dominant quark-core dynamics have been isolated.

More noticeable differences emerge near threshold, where the extrapolation lies below the phenomenological parametrization and available data. This is precisely the region in which the prescription is least reliable and in which final-state interactions, coupled-channel dynamics, resonances, and the complex analytic structure become important.

The comparison with Ref.~\cite{Fu:2025vkq} also suggests that meson-cloud contributions primarily enhance the effective form factor close to the production threshold, whereas at larger momentum transfers the reaction becomes progressively dominated by the dressed-quark core. This interpretation is consistent with the general expectation that long-range hadronic dynamics play their largest role in the low-energy regime, while short-distance quark dynamics dominate at larger $q^2$.

Accordingly, agreement away from threshold tests only the broad energy dependence generated by the extrapolated space-like parametrizations; it does not validate the continuation as a description of the full physical time-like amplitude.


\section{SUMMARY AND CONCLUSIONS}
\label{sec:summary}

We have studied the elastic electromagnetic structure of the $\Omega$ baryon using a symmetry-preserving contact interaction in rainbow--ladder truncation. The calculation combines a Poincar\'e-covariant quark--diquark Faddeev equation with a conserved electromagnetic current whose transverse quark--photon component contains a phenomenological dressed-quark AMM term.

The $\Omega$ baryon, composed exclusively of strange quarks, provides a particularly suitable system for investigating the interplay between quark mass, spin structure, and dynamical chiral symmetry breaking in a
multi-strange environment. Within the present framework, the electromagnetic current is characterized by four independent multipole form factors corresponding to the electric monopole ($G_{E0}$), magnetic dipole ($G_{M1}$), electric quadrupole ($G_{E2}$), and magnetic octupole ($G_{M3}$) moments.

In the space-like region, all four multipole form factors have been computed over a broad range of momentum transfer. The magnetic dipole and higher multipoles display the greatest sensitivity to the AMM parameter. However, the calculated magnetic moment at $\eta=1/6$ is about $36\%$ larger in magnitude than experiment, and increasing positive $\eta$ worsens the discrepancy. This comparison limits the quantitative interpretation of the AMM sensitivity and indicates that additional momentum dependence and/or vertex structures are needed. The $\eta$-variation bands quantify only parameter sensitivity and should not be interpreted as complete theoretical uncertainties.

The large-$Q^2$ behavior of the form factors was analyzed through simple asymptotic parametrizations. We find that the leading multipoles approximately follow a $1/Q^4$ suppression, whereas the higher multipoles are more strongly damped, approximately as $1/Q^6$. These results provide compact representations of the numerical solutions and establish the basis for the analytic continuation into the time-like region.

The fitted space-like results were also continued into the time-like domain using an asymptotically motivated, model-dependent prescription. The resulting effective form factor follows the broad decreasing trend of the data at intermediate and larger $q^2$, whereas the discrepancy near threshold is substantial. Since the continuation produces real functions and contains no resonances, final-state interactions, unitarity cuts, or physical phases, it should not be interpreted as a dynamical description of the time-like amplitude. Its role here is limited to testing how the broad behavior of the space-like parametrizations maps onto the experimentally accessible effective form factor.

The principal outcome of this study is therefore a systematic assessment, within one fixed contact-interaction truncation, of the response of each $\Omega$ multipole to the transverse AMM term. The comparison with the measured magnetic moment provides an important constraint on that response and clearly delineates the quantitative limitations of the framework.

Future work will focus on extending the present approach to momentum-dependent interactions and on incorporating meson-cloud and coupled-channel effects in order to improve the description of the near-threshold time-like region and its associated analytic structure.




\begin{acknowledgements}
L.A. acknowledges financial support from SECIHTI through the program ``Posdoctorados por M\'exico'' under CVU No.~446053.
B. Almeida Zamora acknowledges SECIHTI for the postdoctoral fellowship under CVU No.~935777.
A.B. acknowledges financial support provided by the Beatriz-Galindo programme during his scientific stay at the University of Huelva, Huelva, Spain. 
Otherwise, this work has been partially financed by 
{\em Coordinaci\'on de la Investigaci\'on Cient\'ifica} of the {\em Universidad Michoacana de San Nicol\'as de Hidalgo}, Morelia, Mexico, grant no. 4.10;
{\em Consejo Nacional de Humanidades, Ciencias y Tecnolog\'ias}, Mexico, project no. CBF2023-2024-3544;
Ministerio Español de Ciencia, Innovación y Universidades under grant No. PID2022-140440NB-C22;
Junta de Andalucía under contract Nos. PAIDI FQM-370 and PCI+D+i under the title: ``Tecnologías avanzadas para la exploración del universo y sus componentes" (Code AST22-0001).
\end{acknowledgements}


\bibliography{print_EMFFsOmega}

\end{document}